\documentclass[12pt]{article}
\usepackage{epsfig}
\usepackage{amsfonts}
\usepackage{latexsym}
\usepackage{amsmath}
\usepackage{mathrsfs}
\usepackage{hyperref}
\usepackage{setspace}
\usepackage{color}
\usepackage{bm}
\numberwithin{equation}{section}

\begin{document}

\begin{titlepage}
\unitlength = 1mm
\begin{flushright}
KOBE-COSMO-16-09\\
\end{flushright}

\vskip 1cm
\begin{center}

 {\Large {\textsc{\textbf{Quantum discord in de Sitter space}}}}

\vspace{1.8cm}
Sugumi Kanno$^{*\,\flat}$, Jonathan P. Shock$^{\natural\,\dag}$ and Jiro Soda$^\ddag$

\vspace{1cm}

\shortstack[l]
{\it $^*$ Department of Theoretical Physics and History of Science,
University of the Basque Country\\
~~48080 Bilbao, Spain\\
\it $^\flat$ IKERBASQUE, Basque Foundation for Science, 
Maria Diaz de Haro 3,
48013, Bilbao, Spain\\
\it $^\natural$ Laboratory for Quantum Gravity \& Strings and Astrophysics, Cosmology \& Gravity Center,\\
~~Department of Mathematics \& Applied Mathematics, University of Cape Town,\\
~~Private Bag, Rondebosch 7701, South Africa\\
\it $^\dag$ National Institute for Theoretical Physics,
Private Bag X1,
Matieland, 7602, South Africa\\
\it $^\ddag$ Department of Physics, Kobe University, Kobe 657-8501, Japan}

\vskip 1.5cm

\begin{abstract}
\baselineskip=6mm
We study quantum discord between two free modes of a massive scalar field in a maximally entangled state in de Sitter space. We introduce two observers, one in a global chart and the other in an open chart of de Sitter space, and the observers determine the quantum discord created by each detecting one of the modes. 
This situation is analogous to the relationship between an observer in a Minkowski chart and another in one of the two Rindler charts in flat space. 
We find that the state becomes less entangled as the curvature of the open chart gets larger. In particular, for the cases of a massless, and a conformally coupled scalar field, the entanglement vanishes in the limit of infinite curvature.    
However, we find that the quantum discord never disappears even in the limit that entanglement disappears. 
\end{abstract}

\vspace{1.0cm}

\end{center}
\end{titlepage}

\pagestyle{plain}
\setcounter{page}{1}
\newcounter{bean}
\baselineskip18pt

\setcounter{tocdepth}{2}

\tableofcontents

\section{Introduction}

Quantum entanglement, predicted by Einstein-Podolsky-Rosen (EPR) in 1935 has fascinated many physicists because of its highly counterintuitive properties~\cite{Einstein:1935rr}. For a long time, it was regarded as a problem of philosophy which could not be tested. In 1981, however, Aspect et al. succeeded in showing experimental evidence of quantum entanglement~\cite{Aspect:1981zz,Aspect:1982fx}. 
Since then, more interest has been paid to how to make use of the quantum entanglement of EPR pairs in the fields of quantum cryptography, quantum information and quantum teleportation~\cite{Horodecki:2009zz}.

Quantum entanglement due to pair creation is well studied in non-relativistic regimes. One interesting feature of pair creation in relativistic quantum theory is that of observer dependence~\cite{Garriga:2012qp,Garriga:2013pga,Frob:2014zka}.
For instance, the quantum entanglement between two free modes of a scalar field becomes less entangled if observers who detect each mode are relatively accelerated. In ~\cite{FuentesSchuller:2004xp,Alsing:2006cj}, they considered two free modes of a scalar field in flat space. One is detected by an observer in an inertial frame and the other by a uniformly accelerated observer. They  evaluated the entanglement negativity, a measure of entanglement for mixed states, between the two free modes, which started in a maximally entangled state, to characterize the quantum entanglement and found that the entanglement disappeared for the observer in the limit of infinite acceleration. 

Observer dependent entanglement can also be discussed in the context of an expanding universe. The expansion of the universe produces pair creation. Quantum entanglement has been investigated in this situation in ~\cite{Ball:2005xa,Fuentes:2010dt,Nambu:2011ae}. To see if  entanglement itself exists between causally disconnected regions of de Sitter space,  entanglement entropy, a measure of entanglement for pure states, has been studied in the Bunch-Davies vacuum~\cite{Maldacena:2012xp} and in $\alpha$-vacua~ \cite{Kanno:2014lma,Iizuka:2014rua}. Entanglement negativity has been also studied in \cite{Kanno:2014bma}. Since it was found that the entanglement between causally separated regions in de Sitter space exists, in ~\cite{Kanno:2014ifa, Dimitrakopoulos:2015yva} the question was asked whether there are observable effects of entanglement on the cosmic microwave background (CMB) in our universe.

It was recently shown that quantum entanglement is not the only kind of quantum correlations possible, and are merely a particular characterization of quantumness. In fact, other quantum correlations have now been experimentally found, and quantum discord is known to be a measure of all quantum correlations, including entanglement~\cite{OZ,HV}. This measure can be non-zero even in the absence of entanglement.  
Quantum discord is a useful measure to discuss the performance of quantum computers, which has lead to a lot of works on this topic~\cite{Modi}.
In order to see the observer dependence of all quantum correlations (ie. the total quantumness), the quantum discord between two free modes of a scalar field in flat space, which are detected by two observers in inertial and non-inertial frames respectively, has also been discussed~\cite{Datta,Wang:2009qg}. They found that the quantum discord, in contrast to the entanglement, never disappears, even in the limit of infinite acceleration. 

To construct a theory of gravity compatible with quantum field theory is one of the biggest challenges in modern physics. 
Hence, in trying to understand this problem fully it is important to study quantumness in the context of curved space. Moreover, it is one of the cornerstones of inflationary cosmology that the large scale structure of our universe and temperature fluctuations of the CMB originate from quantum fluctuations during the initial inflationary era. Therefore, this study of quantumness in curved spaces may itself lead to a better understanding of the initial stages of our universe and more precise predictions for cosmological observations. To this aim, quantum discord has recently been studied in a cosmological context~\cite{Lim:2014uea,Martin:2015qta}.

In this paper, we extend the study of quantum discord in Rindler space to that in de Sitter space. In the case of Rindler space, the effect of a non-inertial frame on the quantum correlations has been studied. Since the non-inertial observer in Rindler space corresponds to the observer in an open chart of de Sitter space, we will see the effect of the curvature of the open chart
 on the quantum discord between two free modes of a massive scalar field in de Sitter space.

The organization of the paper is as follows.  In section 2, we review the quantum information theoretic basis of quantum discord.
In section 3, we quantize a free massive scalar field in de Sitter space and express the Bunch-Davies vacuum in a global chart in term of the Fock space in open charts of de Sitter space. 
In section 4, we introduce two observers, Alice and Bob, who detect two free modes of the scalar field which start in an entangle state. We obtain the density operator for Alice and Bob.
In section 5, we compute the entanglement negativity. In section 6, we evaluate the quantum discord in de Sitter space.
In the final section we summarize our results and provide an outlook for possible future calculations.

\section{Quantum discord}
\label{s2}

Quantum discord is a measure of all quantum correlations including entanglement for two subsystems~\cite{OZ,HV}. For a mixed state, this measure can be nonzero even if the state is unentangled. It is defined by quantum mutual information and computed by optimizing over all possible measurements that can be performed on one of the subsystems. 

In classical information theory, the mutual information between two random variables $A$ and $B$ is defined as
\begin{eqnarray}
I(A,B)=H(A)+H(B)-H(A,B)\,,
\label{MI1}
\end{eqnarray}
where the Shannon entropy $H(A)\equiv-\sum_A P(A)\log_2 P(A)$ is used to quantify the ignorance of information about the variable $A$ with probability $P(A)$, and the joint entropy $H(A,B)=-\sum_{A,B}P(A,B)\log_2 P(A,B)$ with the joint probability $P(A,B)$ of both events $A$ and $B$ occuring. The mutual information  (\ref{MI1}) measures how much information $A$ and $B$ have in common.

Using Bayes theorem, the joint probability can be written in terms of the conditional probability as
\begin{equation}
P(A,B)=P(B)P(A|B)\,,
\end{equation} 
where $P(A|B)$ is the probability of $A$ given $B$. The joint entropy can then be written as $H(A,B)=-\sum_{A,B}P(A,B)\left(\log_2 P(B)+\log_2 P(A|B)\right)$. 

Plugging this into Eq.~(\ref{MI1}), the mutual information can be expressed as
\begin{eqnarray}
I(A,B)=H(A)-H(A|B)\,,
\label{MI2}
\end{eqnarray}
where  $P(B)=\sum_A P(A,B)$ has been used, and the conditional entropy is defined as
 $H(A|B)=-\sum_{A,B}P(B)P(A|B)\log_2 P(A|B)$. ie. the average over $B$ of Shannon entropy of event $A$, given $B$. Eqs.~(\ref{MI1}) and (\ref{MI2}) are classically equivalent expressions for the mutual information.

If we try to generalize the concept of mutual information to quantum system, the above two equivalent expressions do not yield identical results because measurements performed on subsystem $B$ disturb subsystem $A$. In  a quantum system, the Shannon entropy is replaced by the von Neumann entropy $S(\rho)=-{\rm Tr}\rho\log_2\rho$ where $\rho$ is a density matrix. The probabilities $P(A,B)$, $P(A)$ and $P(B)$ are replaced respectively by the density matrix of the whole system $\rho_{A,B}$, the reduced density matrix of subsystem $A$,  $\rho_A={\rm Tr}_B\rho_{A,B}$,  and the reduced density matrix of subsystem $B$, 
 $\rho_B={\rm Tr}_A\rho_{A,B}$. 
 
In order to extend the idea of the conditional probability $P(A|B)$ to the quantum system, we use projective measurements of $B$ 
described by a complete set of projectors $\{\Pi_i\}=\{|\psi_i\rangle\langle\psi_i|\}$, where  $i$ distinguishes different outcomes of a measurement on $B$. There are of course many different sets of measurements that we can make. Then the state of $A$ after the measurement of $B$ is given by 
\begin{eqnarray}
\rho_{A|i}=\frac{1}{p_i}{\rm Tr}_B\left(\Pi_i\rho_{A,B}\Pi_i\right)\,,\qquad
p_i={\rm Tr}_{A,B}\left(\Pi_i\rho_{A,B}\Pi_i\right)\,.
\end{eqnarray}
A quantum analog of the conditional entropy can then be defined as 
\begin{eqnarray}
S(A|B)\equiv\min_{\{\Pi_i\}}\sum_ip_i\,S(\rho_{A|i})\,,
\end{eqnarray}
which corresponds to the measurement that least disturbs the overall quantum state, that is, to avoid dependence on the projectors.

Thus, the quantum mutual information corresponding to the two expressions Eqs.~(\ref{MI1}) and ($\ref{MI2}$) is defined respectively by
\begin{eqnarray}
{\cal I}(A,B)&=&S(\rho_A)+S(\rho_B)-S(\rho_{A,B})\,,\\
{\cal J}(A,B)&=&S(\rho_A)-S(A|B)\,.
\end{eqnarray}
The quantum discord is then defined as the difference between the  above two expressions
\begin{eqnarray}
{\cal D}(A,B)&=&{\cal I}(A,B)-{\cal J}(A,B)\nonumber\\
&=&S(\rho_B)-S(\rho_{A,B})+S(A|B)\,.
\end{eqnarray}
The quantum discord thus vanishes in classical mechanics, however it appears not to in some quantum systems. 

\section{From global chart  to open chart}
\label{s3}
Mode functions of a free massive scalar field in open charts of de Sitter space were studied in detail in~\cite{Sasaki:1994yt}. By using the Bogoliubov transformation between the global and open charts, the density matrix and entanglement entropy were calculated in~\cite{Maldacena:2012xp}. In this section, we review their results.

\begin{figure}[t]
\vspace{-3cm}
\includegraphics[height=11cm]{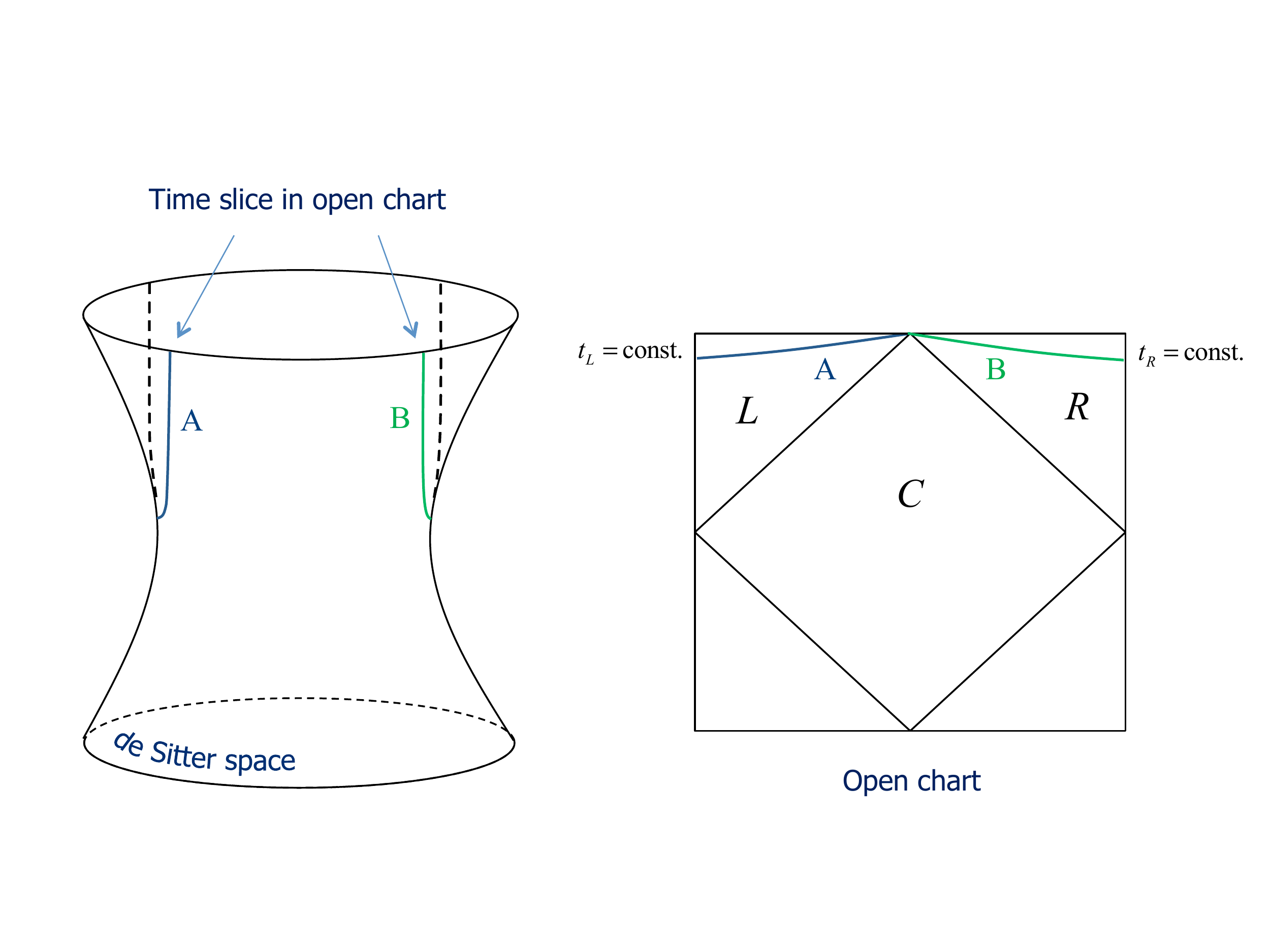}\centering
\vspace{-1cm}
\caption{De Sitter space and its Penrose diagram. The regions $R$ and $L$ are the causally  disconnected open charts of de Sitter space.}
\label{fig1}
\end{figure}

We consider a free scalar field $\phi$ with mass $m$ in de Sitter space represented by the metric $g_{\mu\nu}$.  The action is given by
\begin{eqnarray}
S=\int d^4 x\sqrt{-g}\left[\,-\frac{1}{2}\,g^{\mu\nu}
\partial_\mu\phi\,\partial_\nu \phi
-\frac{m^2}{2}\phi^2\,\right]\,.
\end{eqnarray} 
The coordinate systems of open charts in de Sitter space with the Hubble radius $H^{-1}$ can be obtained by analytic continuation from the Euclidean metric and divided into three parts which we call $R$, $C$ and $L$ as shown in Figure~\ref{fig1}. Their metrics are given respectively by
\begin{eqnarray}
ds^2_R&=&H^{-2}\left[-dt^2_R+\sinh^2t_R\left(dr^2_R+\sinh^2r_R\,d\Omega^2\right)
\right]\,,\nonumber\\
ds^2_C&=&H^{-2}\left[dt_C^2+\cos^2t_C\left(-dr_C^2+\cosh^2r_C\,d\Omega^2\right)\right]\,,\nonumber\\
ds^2_L&=&H^{-2}\left[-dt^2_L+\sinh^2t_L\left(dr^2_L+\sinh^2r_L\,d\Omega^2\right)
\right]\,,
\end{eqnarray}
where $d\Omega^2$ is the metric on the two-sphere. Note that the region $L$ and $R$ covered by the coordinates $(t_L, r_L)$ and $(t_R, r_R)$ respectively are the two causally  disconnected open charts of de Sitter space\footnote{The point between $R$ and $L$ regions is a part of the timelike infinity where infinite volume exists.}. The region $C$ is covered by the coordinate $(r_C, t_C)$.

The solutions of the Klein-Gordon equation are expressed by
\begin{eqnarray}
u_{\sigma p\ell m}(t,r,\Omega)\sim\frac{H}{\sinh t}\,
\chi_{p,\sigma}(t)\,Y_{p\ell m} (r,\Omega)\,,\qquad
-{\rm\bf L^2}Y_{p\ell m}=\left(1+p^2\right)Y_{p\ell m}\,,
\end{eqnarray}
where $(t,r)=(t_R,r_R)$ or $(t_L,r_L)$ and $Y_{p\ell m}$ are harmonic functions on the three-dimensional hyperbolic space.
The eigenvalues $p$ normalized by $H$ take positive real values. 
 The positive frequency mode functions corresponding to the Euclidean vacuum (the Bunch-Davies vacuum) that are supported both on the $R$ and $L$ regions are
\begin{eqnarray}
\chi_{p,\sigma}(t)=\left\{
\begin{array}{l}
\frac{e^{\pi p}-i\sigma e^{-i\pi\nu}}{\Gamma(\nu+ip+\frac{1}{2})}P_{\nu-\frac{1}{2}}^{ip}(\cosh t_R)
-\frac{e^{-\pi p}-i\sigma e^{-i\pi\nu}}{\Gamma(\nu-ip+\frac{1}{2})}P_{\nu-\frac{1}{2}}^{-ip}(\cosh t_R)
\,,\\
\\
\frac{\sigma e^{\pi p}-i\,e^{-i\pi\nu}}{\Gamma(\nu+ip+\frac{1}{2})}P_{\nu-\frac{1}{2}}^{ip}(\cosh t_L)
-\frac{\sigma e^{-\pi p}-i\,e^{-i\pi\nu}}{\Gamma(\nu-ip+\frac{1}{2})}P_{\nu-\frac{1}{2}}^{-ip}(\cosh t_L)
\,,
\label{solutions}
\end{array}
\right.
\end{eqnarray}
where $P^{\pm ip}_{\nu-\frac{1}{2}}$ are the associated Legendre functions
and the index $\sigma$ takes the values $\pm 1$ which distinguishes two 
independent solutions for each region~\cite{Sasaki:1994yt}. Note that the effect of the curvature of the three-dimensional hyperbolic space starts to appear around $p\sim 1$. And the effects gets stronger as $p$ becomes smaller than $1$. Thus we can probe the effect of the curvature on quantum entanglement by varying $p$.  We define a mass parameter
\begin{eqnarray}
\nu=\sqrt{\frac{9}{4}-\frac{m^2}{H^2}}\,.
\end{eqnarray}
The normalization factor for the solutions in Eq.~(\ref{solutions}) is given by
\begin{eqnarray}
N_{p}=\frac{4\sinh\pi p\,\sqrt{\cosh\pi p-\sigma\sin\pi\nu}}{\sqrt{\pi}\,|\Gamma(\nu+ip+\frac{1}{2})|}\,.
\label{norm}
\end{eqnarray}

Since they form a complete orthonormal set of modes, the field can be expanded in terms of the creation and annihilation operators
\begin{eqnarray}
\hat\phi(t,r,\Omega) = \int dp \sum_{\sigma,\ell,m} 
\left[\,a_{\sigma p\ell m}\,u_{\sigma p\ell m}(t,r,\Omega)
+a_{\sigma p\ell m}^\dagger\,u^*_{\sigma p\ell m}(t,r,\Omega)\,\right]\,,
\end{eqnarray}
where $a_{\sigma p\ell m}$ satisfies $a_{\sigma p\ell m}|0\rangle_{\rm BD}=0$ and the commutation relations are $[a_{\sigma p\ell m},a_{\sigma' p'\ell' m'}^\dag]=\delta(p-p')\delta_{\sigma,\sigma'}\delta_{\ell,\ell'}\delta_{m,m'}$.
In the following the indices $p, \ell, m$ of the operators and mode functions are omitted for simplicity unless there may be any confusion.

The Bogoliubov transformation between the Bunch-Davies vacuum and $R$, $L$ vacua, 
derived in~\cite{Maldacena:2012xp}, is expressed as
\begin{eqnarray}
|0\rangle_{\rm BD}=\sqrt{1-|\gamma_p|^2}\,e^{\gamma_p\,c_R^\dag c_L^\dag}\,
|0\rangle_R|0\rangle_L\,,
\label{bogoliubov1}
\end{eqnarray}
where $|0\rangle_R, |0\rangle_L$ are annihilated by $c_R$, $c_L$ respectively and
\begin{eqnarray}
\gamma_p= i\frac{\sqrt{2}}{\sqrt{\cosh 2\pi p + \cos 2\pi \nu}
 + \sqrt{\cosh 2\pi p + \cos 2\pi \nu +2 }}\,.
\label{gammap}
\end{eqnarray}
Note that in the case of conformal invariance ($\nu=1/2$) and masslessness ($\nu=3/2$), we find $|\gamma_p|=e^{-\pi p}$. 

The relation of creation and annihilation operators between the Bunch-Davies vacuum and $R$, $L$ vacua is given by
\begin{eqnarray}
a_\sigma^\dag&=&\frac{N_p}{N_b}\left[\,
\bar\gamma_{R\sigma}\,e^{-i\frac{\theta}{2}}
\left(u\,c_R^\dag-v\,c_R\right)
+\delta_{R\sigma}\,e^{i\frac{\theta}{2}}
\left(u\,c_R-v\,c_R^\dag\right)\right.\nonumber\\
&&\left.\qquad+\,
\bar\gamma_{L\sigma}\,e^{-i\frac{\theta}{2}}
\left(u\,c_L^\dag-v\,c_L\right)
+\delta_{L\sigma}\,e^{i\frac{\theta}{2}}
\left(u\,c_L-v\,c_L^\dag\right)
\right]      \ , 
\label{ac}
\end{eqnarray}
where $N_p/N_b$ is the ratio of the normalizations of the mode functions in the Bunch-Davies vacuum Eq.~(\ref{norm}) and in the $R$, $L$ vacua
\begin{eqnarray}
\frac{N_p}{N_b}&=&\frac{4\sinh\pi p\sqrt{\cosh\pi p-\sigma\sin\pi\nu}}{\sqrt{2\pi p}}\frac{|\Gamma(1+ip)|}{|\Gamma(\nu+ip+\frac{1}{2})|} \,.
\end{eqnarray}
The phase factor that appears in the Bogoliubov transformation is
\begin{eqnarray}
e^{i\theta}=-\frac{\Gamma\left(\nu-ip+\frac{1}{2}\right)}{\Gamma\left(\nu+ip+\frac{1}{2}\right)}\frac{\cosh\left(\pi p+i\pi\nu\right)}{\sqrt{\sinh^2\pi p+\cos^2\pi\nu}}\,,
\label{phase}
\end{eqnarray}
and other variables which originally come from the coefficients of the Legendre functions in Eq.~(\ref{solutions}) are
\begin{eqnarray}
\bar{\gamma}_{R\sigma}&=&\sigma\,\bar{\gamma}_{L\sigma}=\frac{\Gamma(\nu-ip+\frac{1}{2})}{8\sinh\pi p}\,\frac{e^{\pi p}-i\sigma e^{-i\pi\nu}}{\cosh\pi p-\sigma\sin\pi\nu}\nonumber  \ , \\
\delta_{R\sigma}&=&\sigma\,\delta_{L\sigma}=\frac{\Gamma(\nu+ip+\frac{1}{2})}{8\sinh\pi p}\,\frac{e^{-\pi p}-i\sigma e^{-i\pi\nu}}{\cosh\pi p-\sigma\sin\pi\nu} \,,
\end{eqnarray}
where $\sigma=\pm 1$. $u$ and $v$ are given by
\begin{eqnarray}
u&=&\frac{1-\gamma_k\,\zeta}{\sqrt{\left(1-\gamma_k\,\zeta\right)^2-\omega^2}}\,,\qquad
v=\frac{-\omega}{\sqrt{\left(1-\gamma_k\,\zeta\right)^2-\omega^2}} \,,
\label{uv}
\end{eqnarray}
where
\begin{eqnarray}
\omega=\frac{\sqrt{2}\,e^{-\pi k}}{\sqrt{\cosh 2\pi k+\cos 2\pi\nu}}\cos\pi\nu\,,\qquad
\zeta=i\frac{\sqrt{2}\,e^{-\pi k}}{\sqrt{\cosh 2\pi k+\cos 2\pi\nu}}\sinh\pi k \,.
\label{omegazeta}
\end{eqnarray}

\section{Setup of quantum states}
\label{s4}

The solutions in the Bunch-Davies vacuum~(\ref{solutions}) are related to those in the $R$, $L$ vacua through Bogoliubov transformations (\ref{bogoliubov1}). Using this transformation, we find that the ground state of a given mode seen by an observer in the global chart corresponds to a two-mode squeezed state in the open charts. These two modes correspond to the fields observed in the $R$ and $L$ charts. If we probe only one of the open charts, say $L$, we have no access to the modes in the causally disconnected $R$ region and must therefore trace over the inaccessible region. This situation is analogous to the relationship between an observer in a Minkowski chart and another in one of the two Rindler charts in flat space, in the sense that the global chart and Minkowski chart cover the whole spacetime geometry while open charts and Rindler charts cover only a portion of the spacetime geometry and thus there exists horizons.

We start with two maximally entangled modes with $p=k$ and $s$ of the free massive scalar field in de Sitter space, 
\begin{eqnarray}
|\psi\rangle=\frac{1}{\sqrt{2}}\,\Bigl(\,|0_s\rangle_{\rm BD}|0_k\rangle_{\rm BD}+|1_s\rangle_{\rm BD}|1_k\rangle_{\rm BD}\,\Bigr)\,.
\label{entangle}
\end{eqnarray}
We assume that Alice has a detector which only detects mode $s$ and Bob has a detector sensitive only to mode $k$. When Bob resides in the $L$ region, the Bunch-Davies vacuum with mode $k$ can be expressed as a two-mode squeezed state of the $R$ and $L$ vacua
\begin{eqnarray}
|0_k\rangle_{\rm BD}=\sqrt{1-|\gamma_k|^2}\,\sum_{n=0}^\infty\gamma_k^n|n_k\rangle_L|n_k\rangle_R\,,
\label{bogoliubov2}
\end{eqnarray}
where we expanded the exponent in Eq.~(\ref{bogoliubov1}).
The $|n_k\rangle_L$ and $|n_k\rangle_R$ refer to the two modes of the $L$ and $R$ open charts. The single particle excitation state is then calculated by operating Eq.~(\ref{ac}) on Eq.~(\ref{bogoliubov2}),
\begin{eqnarray}
|1_k\rangle_{\rm BD}&=&\frac{1-|\gamma_k|^2}{\sqrt{2}}
\sum_{n=0}^\infty\gamma_k^n\sqrt{n+1}\,
\Bigl[\,|(n+1)_k\rangle_{L}\,|n_k\rangle_{R}
+|n_k\rangle_{L}\,|(n+1)_k\rangle_{R}\,\Bigr]\,.
\label{excitation}
\end{eqnarray}
Note that the overall phase factor stemming from Eq.~(\ref{phase}) can be absorbed in the definition of states.

Since the $R$ region is inaccessible to Bob, we need to trace over the states in the $R$ region. If we plug Eqs.~(\ref{bogoliubov2}) and (\ref{excitation}) into the initial maximally entangled state (\ref{entangle}), the reduced density matrix after tracing out the states in the $R$ region is represented as
\begin{eqnarray}
\rho_{A,B}=\frac{1-|\gamma_k|^2}{2}\sum^\infty_{m=0}
|\gamma_k|^{2m}\,\rho_m\,,
\label{rhoAB1}
\end{eqnarray}
where
\begin{eqnarray}
\rho_m&=&|0m\rangle\langle 0m|
+\sqrt{\frac{1-|\gamma_k|^2}{2}}\sqrt{m+1}\,
\Bigl(\,\gamma_k|0\,m+1\rangle\langle 1m|+\gamma_k^*\,|1m\rangle\langle 0\,m+1|
\,\Bigr)
\nonumber\\
&&
+\sqrt{\frac{1-|\gamma_k|^2}{2}}\sqrt{m+1}\,
\Bigl(|0m\rangle\langle 1\,m+1|
+|1\,m+1\rangle\langle 0m|
\,\Bigr)
\nonumber\\
&&+\frac{1-|\gamma_k|^2}{2}\,(m+1)\,\Bigl(|1m\rangle\langle 1m|
+|1\,m+1\rangle\langle 1\,m+1|
\,\Bigr)
\nonumber\\
&&+\frac{1-|\gamma_k|^2}{2}\,\sqrt{(m+1)(m+2)}
\,\Bigl(\gamma_k\,|1\,m+2\,\rangle\langle 1m|
+\gamma_k^*\,|1m\rangle\langle 1\,m+2\,|
\,\Bigl)\,,
\nonumber
\end{eqnarray}
with $|mn\rangle=|m_s\rangle_{\rm BD}|n_k\rangle_{L}$. This is a mixed state. 

\section{Entanglement negativity}
\label{s5}

\begin{figure}[t]
\begin{center}
\vspace{-1cm}
\begin{minipage}{8.1cm}
\includegraphics[height=6cm]{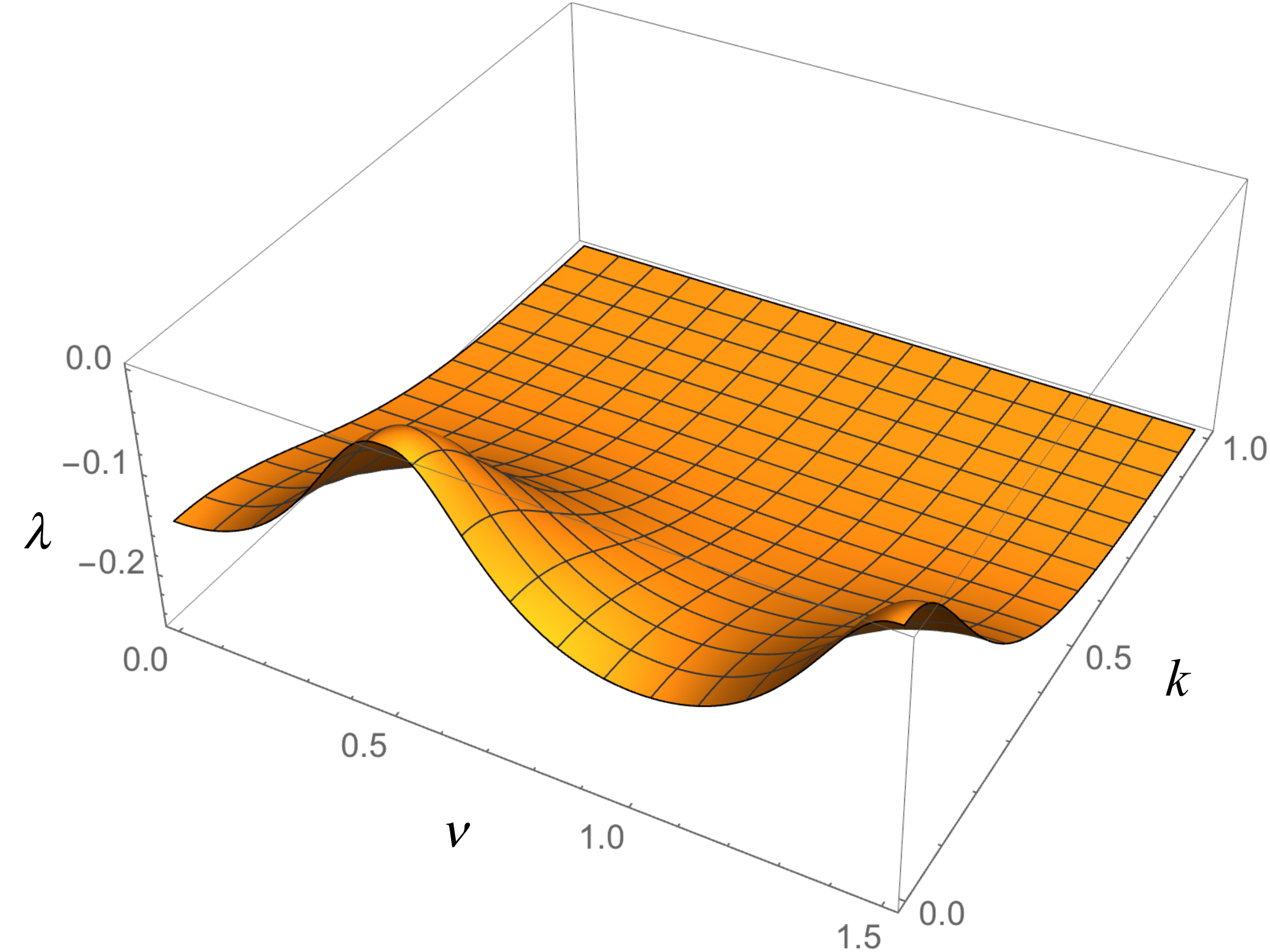}\centering
\end{minipage}
\begin{minipage}{8.1cm}
\hspace{0.5cm}
\includegraphics[height=6cm]{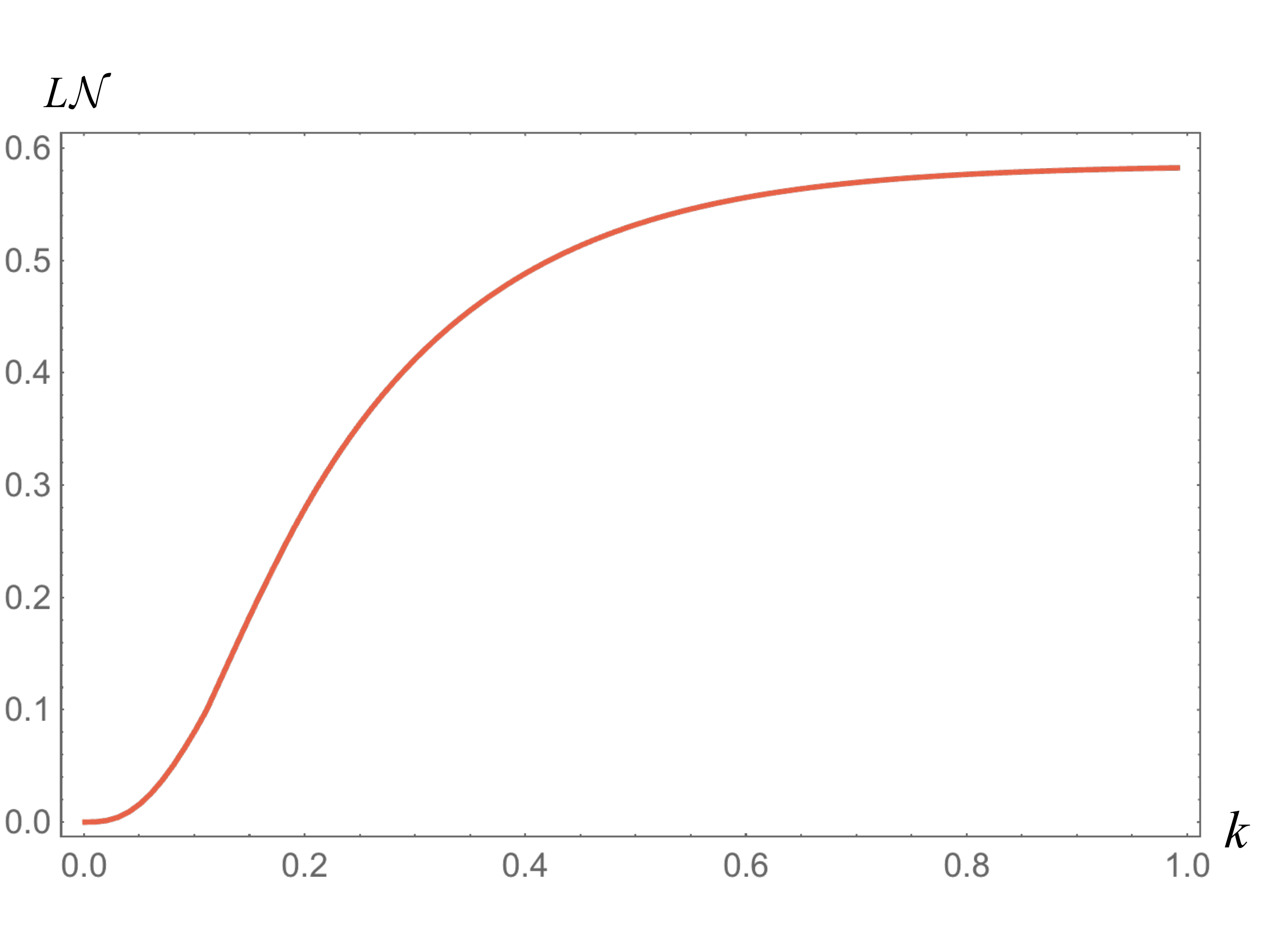}
\end{minipage}
\caption{3D plot of the negative eigenvalues $\lambda$ as a function of $\nu$ and $k$ (Left). The logarithmic negativity vanishes only in the limit of $k\rightarrow 0$ for both $\nu=1/2$ and $3/2$. (Right).}
\label{fig2}
\end{center}
\end{figure}

Let us calcuate the entanglement negativity, a measure of entanglement for mixed states, first. This measure is defined by the partial transpose, the non-vanishing of which provides a sufficient criterion for entanglement. If at least one eigenvalue of the partial transpose is negative, then the density matrix is entangled. 

If we take the partial transpose with respect to Alice's subsystem, then we find
\begin{eqnarray}
\rho_{A,B}^{T}=\frac{1-|\gamma_k|^2}{2}\sum^\infty_{m=0}
|\gamma_k|^{2m}\,\rho_m^{T}\,,
\end{eqnarray}
where
\begin{eqnarray}
\rho_m^T&=&|0m\rangle\langle 0m|
+\sqrt{\frac{1-|\gamma_k|^2}{2}}\sqrt{m+1}\,
\Bigl(\,\gamma_k|1\,m+1\rangle\langle 0m|+\gamma_k^*\,|0m\rangle\langle 1\,m+1|
\,\Bigr)
\nonumber\\
&&
+\sqrt{\frac{1-|\gamma_k|^2}{2}}\sqrt{m+1}\,
\Bigl(|1m\rangle\langle 0\,m+1|+|0\,m+1\rangle\langle 1m|
\,\Bigr)
\nonumber\\
&&+\frac{1-|\gamma_k|^2}{2}\,(m+1)\,\Bigl(|1m\rangle\langle 1m|
+|1\,m+1\rangle\langle 1\,m+1|
\,\Bigr)
\nonumber\\
&&+\frac{1-|\gamma_k|^2}{2}\,\sqrt{(m+1)(m+2)}
\,\Bigl(\gamma_k\,|1\,m+2\,\rangle\langle 1m|
+\gamma_k^*\,|1m\rangle\langle 1\,m+2\,|
\,\Bigl)\,.
\nonumber
\end{eqnarray}
We compute the eigenvalues $\lambda$ numerically and the resultant negative eigenvalues are plotted in the left panel of Figure~\ref{fig2}. We see that the negative eigenvalues for $\nu=1/2$ and $3/2$ go to zero in the limit of $k\rightarrow 0.$\footnote{Here, we use different mode functions from those discussed in~\cite{Kanno:2014bma}. However, the normalization of the mode functions in ~\cite{Kanno:2014bma} had an error. Thus, the resultant Figure~3 in \cite{Kanno:2014bma} is incorrect.}
  In order to see if the entanglement vanishes clearly, we calculate the logarithmic negativity next.

\begin{figure}[t]
\begin{center}
\vspace{-1cm}
\begin{minipage}{8.1cm}
\includegraphics[height=6cm]{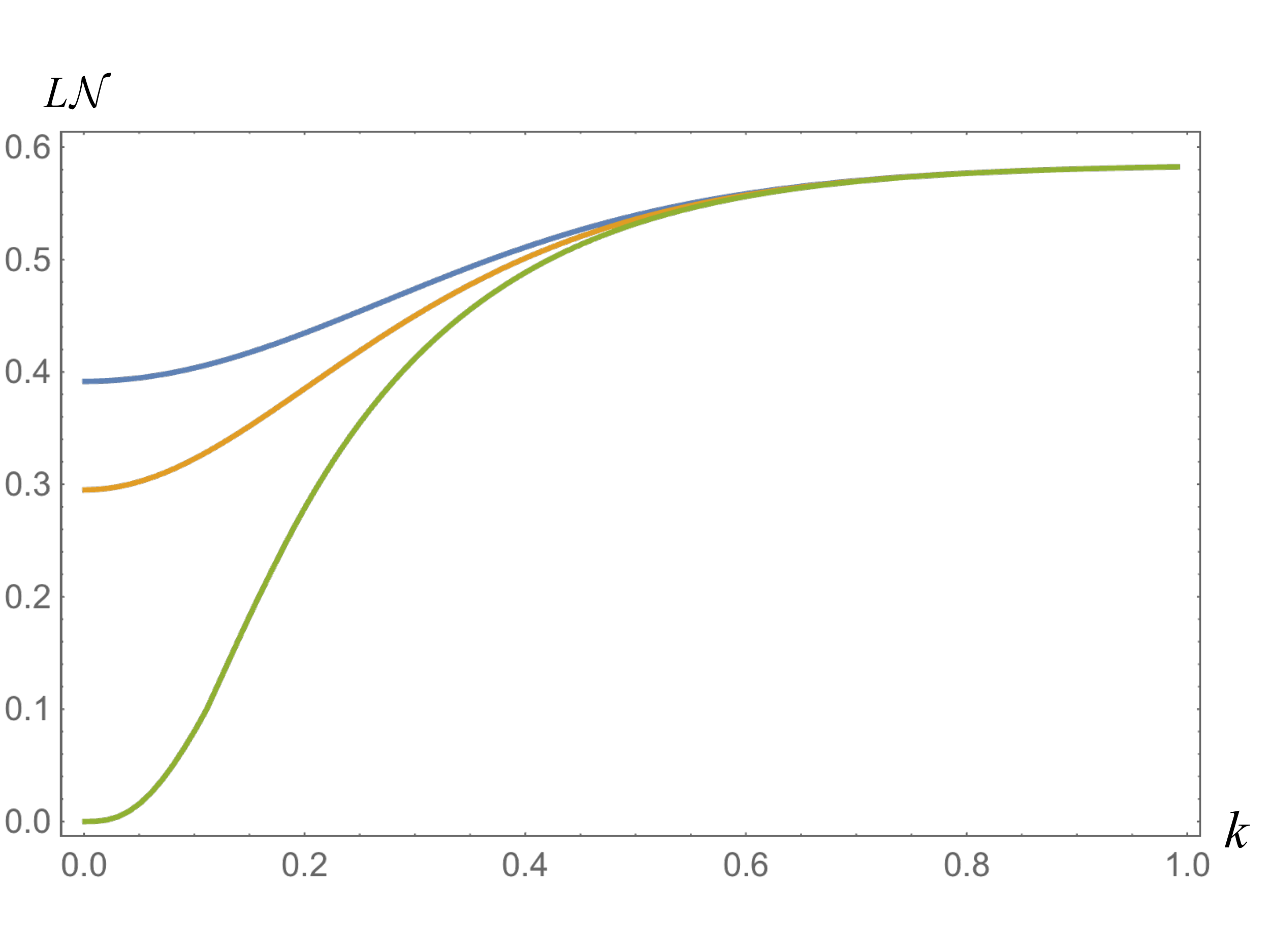}\centering
\end{minipage}
\begin{minipage}{8.1cm}
\hspace{0.5cm}
\includegraphics[height=6cm]{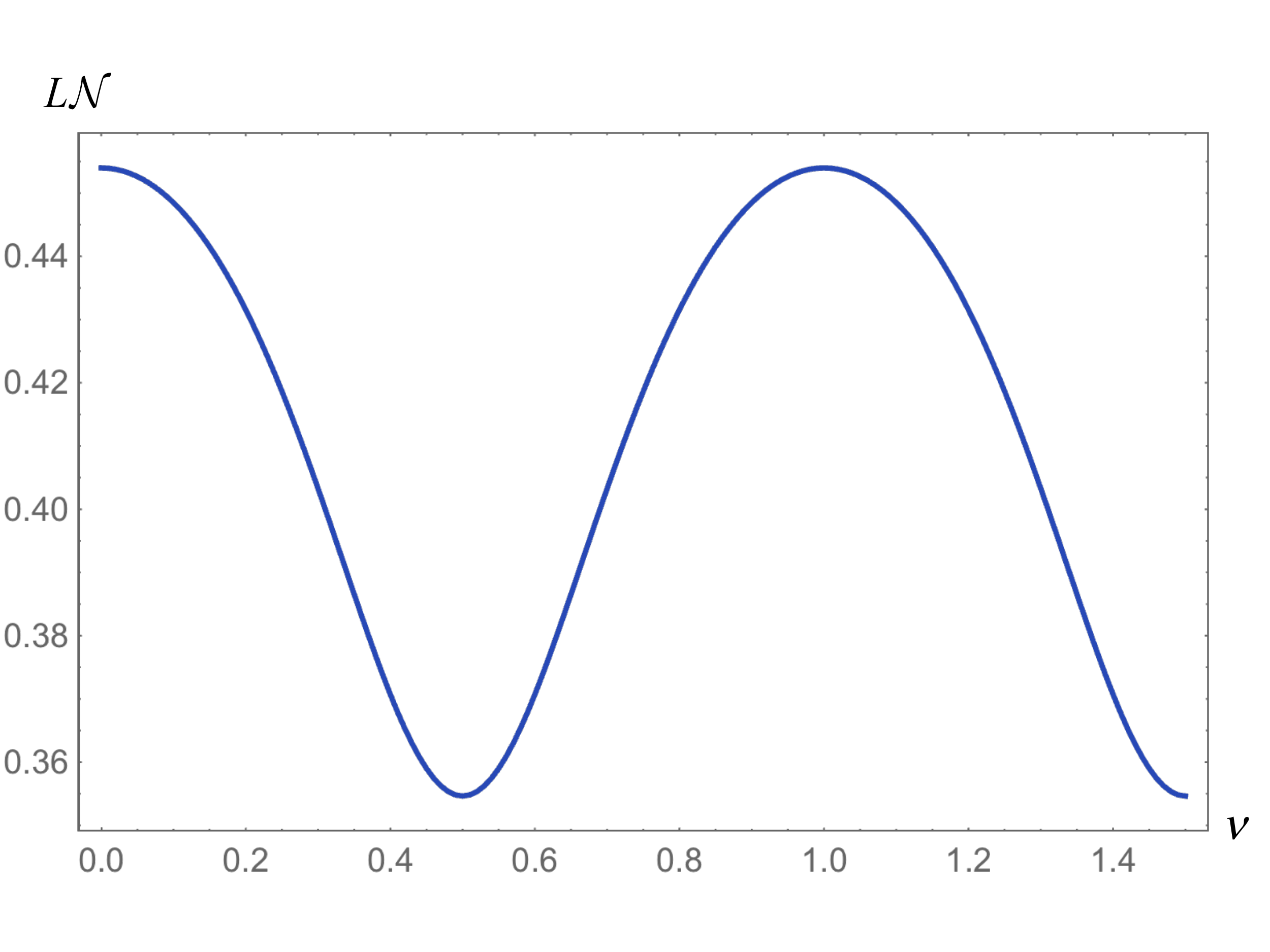}
\end{minipage}
\caption{Plots of the logarighmic negativity as a function of $k$. The Blue line is for $\nu=0$, $1$, the yellow is for $\nu=1/4$, $5/4$ and the green is for $\nu=1/2$, $3/2$. (Left) Plot of the logarithmic negativity as a function of $\nu$ for $k=0.25$. (Right)}
\label{fig3}
\end{center}
\end{figure}

The negativity is defined by summing over all the negative eigenvalues
\begin{eqnarray}
{\cal N}=\sum_{\lambda_i<0}|\lambda_i|\,.
\end{eqnarray}
and there exists no entanglement when ${\cal N}=0$. However, this measure is not additive and not suitable for multiple subsystems. The logarithmic negativity is thus a better measure than the negativity in this scenario and is defined as
\begin{eqnarray}
L{\cal N}=\log_2\left(2{\cal N}+1\right)\,.
\end{eqnarray} 
The state is entangled when $L{\cal N}\neq 0$. We sum over all negative eigenvalues and calculate the logarithmic negativity. 
We focus on Alice and Bob's detectors for modes of momentum $s$ and $k$, and thus don't integrate either over $k$, nor a volume integral over the hyperboloid. We plot the logarithmic negativity in the right panel of Figure~\ref{fig2}. For $\nu=1/2$ and $3/2$, the logarithmic negativity vanishes in the limit of $k\rightarrow 0$, that is, in the limit of infinite curvature. This is consistent with the flat space result where the entanglement vanishes in the limit of infinite acceleration of the observer. Here we stress that the entanglement weakens, but survives even in the limit of infinite curvature for a massive scalar field other than $\nu=1/2$ and $\nu=3/2$ as  can be seen from the left panel of Figure~\ref{fig3}.
The logarithmic negativity as a function of the mass parameter $\nu$ is plotted in Figure~\ref{fig3} where we take $k=0.25$. The qualitative feature of it is similar to the left panel of Figure~\ref{fig2} and we can read off that the entanglement gets smaller for $\nu=1/2$ and $3/2$. The oscillatory behavior comes from the factor
  $\cos2\pi\nu$ in $\gamma_k$ of Eq.~(\ref{gammap}) $(p=k)$.

\section{Quantum discord in de Sitter space}
\label{s6}

Next we calculate the overall quantumness of the system given by the quantum discord.
We will make our measurement on the Alice's side. The quantum discord is
\begin{eqnarray}
{\cal D}_\theta&=&S(\rho_A)-S(\rho_{A,B})+S(B|A)\,.
\label{discord1}
\end{eqnarray}
To calculate the above, we rewrite the state (\ref{rhoAB1}) as
\begin{eqnarray}
\rho_{A,B}=\frac{1-|\gamma_k|^2}{2}\Bigl(\,|0\rangle\langle 0|\otimes M_{00}
+|0\rangle\langle 1|\otimes M_{01}
+|1\rangle\langle 0|\otimes M_{10}
+|1\rangle\langle 1|\otimes M_{11}\,\Bigr)\,,
\label{rhoAB2}
\end{eqnarray}
where we define
\begin{eqnarray}
M_{00}&=&\sum^\infty_{m=0}|\gamma_k|^{2m}\,|m\rangle\langle m|\,,\\
M_{01}&=&\sqrt{\frac{1-|\gamma_k|^2}{2}}\sum^\infty_{m=0}\sqrt{m+1}\,|\gamma_k|^{2m}
\Bigl(\gamma_k\,|m+1\rangle\langle m|
+\,|m\rangle\langle m+1|
\,\Bigr)\,,\\
M_{10}&=&M_{01}^\dag\nonumber\\
&=&\sqrt{\frac{1-|\gamma_k|^2}{2}}\sum^\infty_{m=0}\sqrt{m+1}\,|\gamma_k|^{2m}
\Bigl(\gamma_k^*\,|m\rangle\langle m+1|
+|m+1\rangle\langle m|
\,\Bigr)\,,\\
M_{11}&=&\frac{1-|\gamma_k|^2}{2}\sum^\infty_{m=0}\left(m+1\right)\,|\gamma_k|^{2m}\Bigl(\,|m\rangle\langle m|+|m+1\rangle\langle m+1|
\,\Bigr)\nonumber\\
&&+\frac{1-|\gamma_k|^2}{2}\sum^\infty_{m=0}
\sqrt{\left(m+1\right)\left(m+2\right)}\,|\gamma_k|^{2m}
\Bigl(\gamma_k\,|m+2\rangle\langle m|+\gamma_k^*\,|m\rangle\langle m+2|\,\Bigr)\,.
\end{eqnarray}
The state is split into Alice's subsystem (two dimensions) and Bob's subsystem (infinite dimensional). 
Then Alice's density matrix is easy to obtain as
\begin{eqnarray}
\rho_A={\rm Tr}_B\rho_{AB}
=\frac{1}{2}\Bigl(\,|0\rangle\langle 0|+|1\rangle\langle 1|\,\Bigr)\,,
\end{eqnarray}
and we get $S(\rho_A)=-\rm{Tr}\rho_A\log_2\rho_A=1$. For the von Neumann entropy of the whole system $S(\rho_{A,B})$, we need to find the eigenvalues of $\rho_{A,B}$ numerically.

In order to calculate the quantum conditional entropy $S(B|A)$, we restrict ourselves to projective measurements on Alice's subsystem described by a complete set of projectors
\begin{eqnarray}
\Pi_{\pm}=\frac{I\pm{\bm x}\cdot{\bm\sigma}}{2}
=\frac{1}{2}\left(
\begin{array}{cc}
1\pm x_3 & \pm\left(x_1-ix_2\right)\\
\pm\left(x_1+ix_2\right) & 1\mp x_3
\end{array}
\right)\,,
\end{eqnarray}
where ${\bm x}\cdot{\bm x}=x_1^2+x_2^2+x_3^2=1$, $I$ is the $2\times 2$ identity matrix and ${\bm\sigma}$ are the Pauli matrices. Here, a choice of the $x_i$ corresponds to a choice of measurement, and we will thus be interested in the particular measurement which minimises the disturbance on the system. Then the density matrix after the measurement is
\begin{eqnarray}
\rho_{B|\pm}=\frac{1}{p_\pm}\Pi_\pm\rho_{AB}\Pi_\pm\,.
\end{eqnarray}
The trace of it is calculated as
\begin{eqnarray}
{\rm Tr}\rho_{B|\pm}&=&\frac{1}{p_\pm}{\rm Tr}\left(\Pi_\pm\rho_{AB}\right)
\\
&=&\frac{1-|\gamma_k|^2}{4p_\pm}\Bigl[\left(1\pm x_3\right)M_{00}+\left(1\mp x_3\right)M_{11}
\pm\left(x_1+ix_2\right)M_{01}\pm\left(x_1-ix_2\right)M_{10}\Bigr]\,,\nonumber
\end{eqnarray}
where we used $\Pi_\pm^2=\Pi_\pm$ and
\begin{eqnarray}
p_\pm={\rm Tr}_{AB}\,\Pi_{\pm}\rho_{AB}=\frac{1-|\gamma_k|^2}{4}\Bigr[
\left(1\pm x_3\right){\rm Tr}M_{00}+\left(1\mp x_3\right){\rm Tr}M_{11}
\Bigl]=\frac{1}{2}\,.
\end{eqnarray}
By using the parametrization
\begin{eqnarray}
x_1=\sin\theta\cos\phi\,,\qquad
x_2=\sin\theta\sin\phi\,,\qquad
x_3=\cos\theta\,,
\end{eqnarray}
$\rho_{B|\pm}$ is found to be independent of the phase factor $\phi$:
\begin{eqnarray}
\rho_{B|\pm}=\frac{1-|\gamma_k|^2}{2}\Bigl[
\left(1\pm\cos\theta\right)M_{00}+\left(1\mp\cos\theta\right)M_{11}
\pm\sin\theta M_{01}\pm\sin\theta M_{10}
\Bigr]\,.
\end{eqnarray}

\begin{figure}[t]
\begin{center}
\vspace{-1cm}
\begin{minipage}{8.1cm}
\includegraphics[height=6cm]{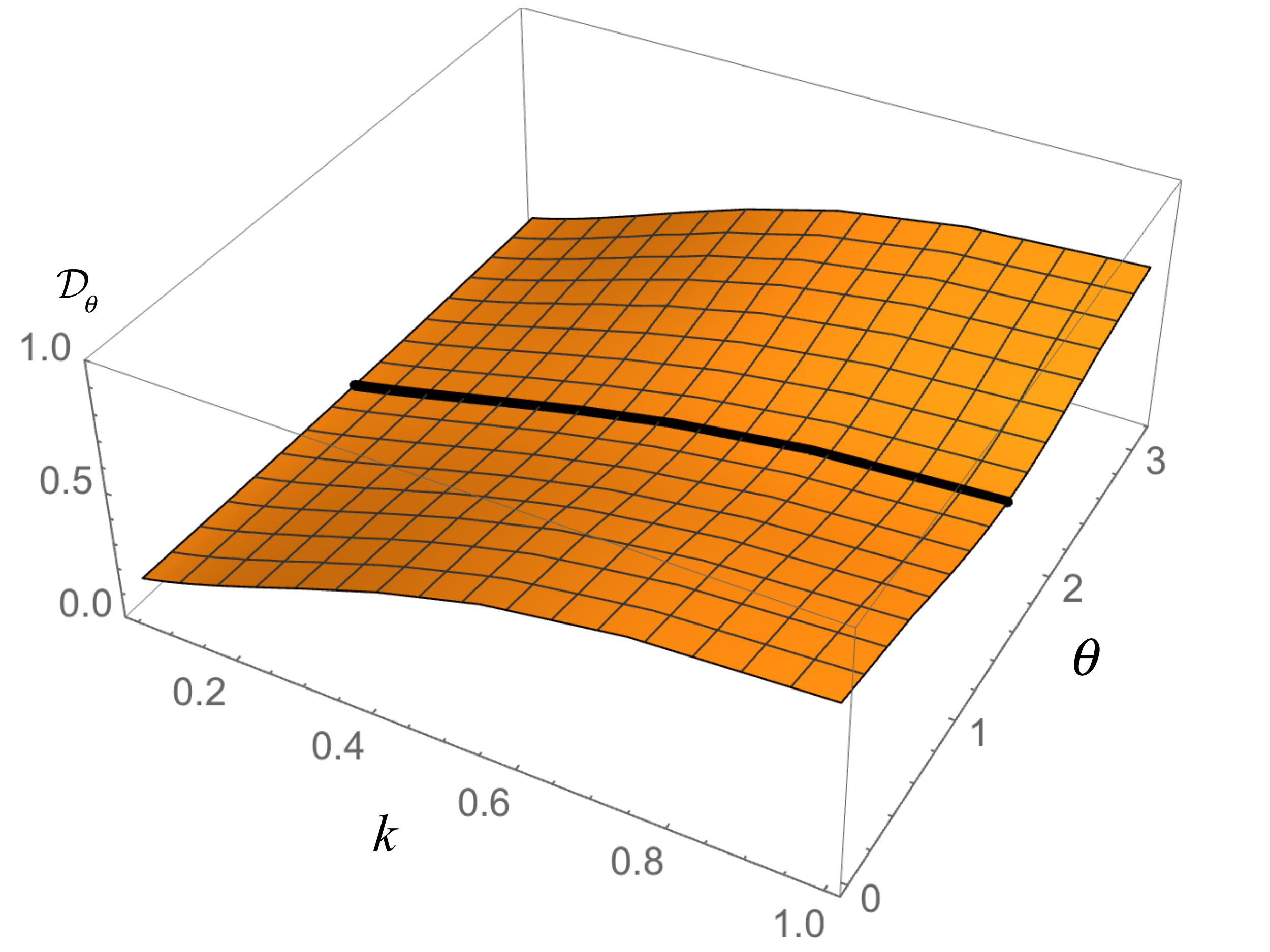}\centering
\end{minipage}
\begin{minipage}{8.1cm}
\hspace{0.5cm}
\includegraphics[height=6cm]{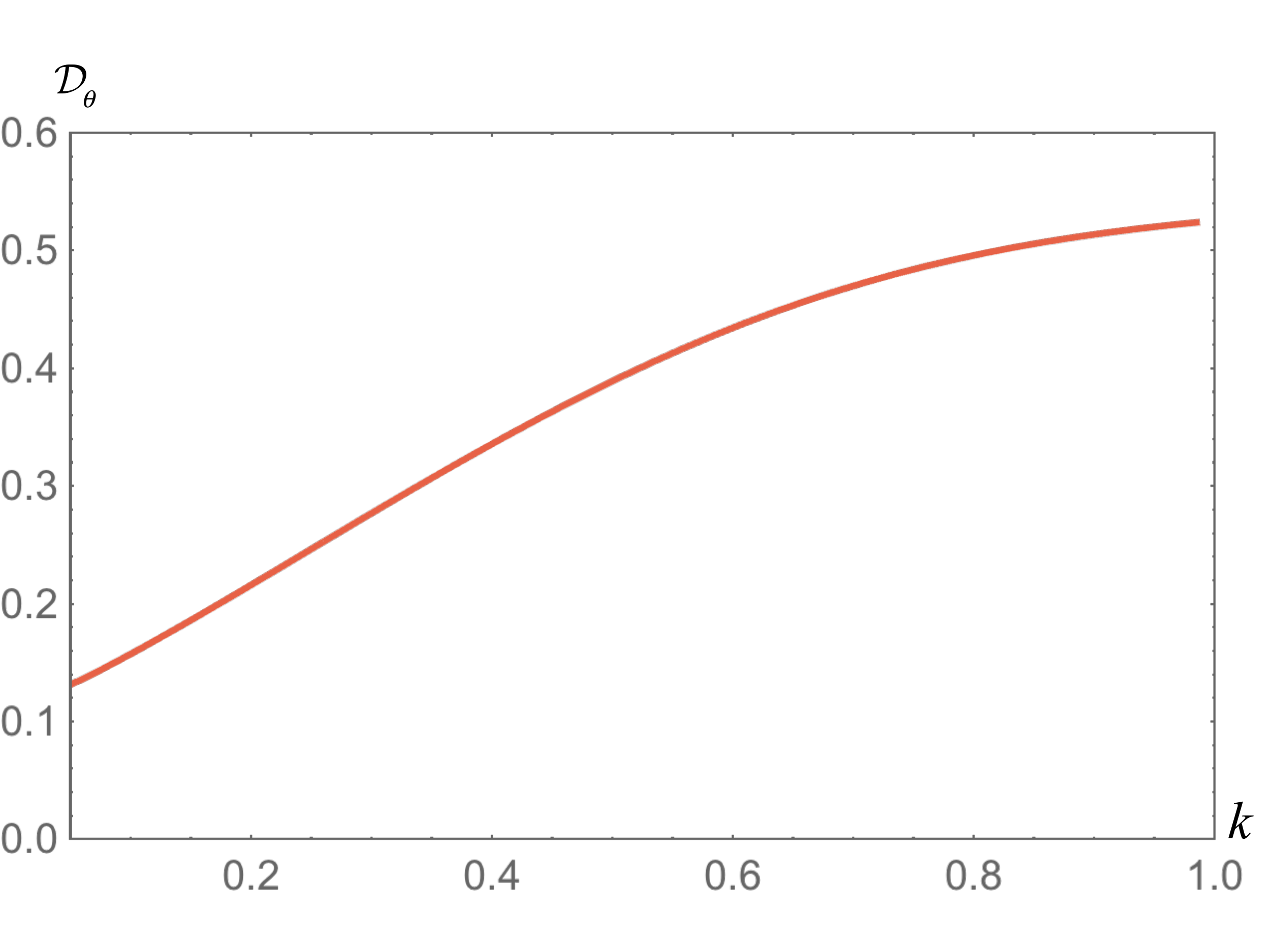}
\end{minipage}
\caption{The left panel shows the 3D plot of the quantum discord as a function of $k$ and $\theta$ for $\nu=1/2$. The black line shows the minima for different values of $k$. The right panel shows the quantum discord has a finite value even in a limit that entanglement vanishes $(k\rightarrow 0)$ for $\nu=1/2$.}
\label{fig4}
\end{center}
\end{figure}

Thus, the quantum discord (\ref{discord1}) is now expressed as
\begin{eqnarray}
{\cal D}_\theta=1+\rm{Tr}\rho_{AB}\log_2\rho_{AB}-\frac{1}{2}\left(\rm{Tr}\rho_{B|+}\log_2\rho_{B|+}+\rm{Tr}\rho_{B|-}\log_2\rho_{B|-}\right)\,.
\label{discord2}
\end{eqnarray}
We will find eigenvalues of $\rho_{A,B}$, $\rho_{B|+}$ and $\rho_{B|-}$ numerically and find $\theta$ that minimizes the above ${\cal D}_\theta$. The left panel of Figure~\ref{fig4} displays the result where the black line shows the minima of ${\cal D}_\theta$ for different values of $k$ and we can read off that $\theta = \pi/2$ minimize ${\cal D}_\theta$. The right panel shows that ${\cal D}_\theta$ has a finite value even in a limit, $k\rightarrow 0$, in which the entanglement vanishes. Note that the convergence of the sum for $\rho_{A,B}$ is not fast for $\nu=1/2$ and $3/2$ in the limit of $k\rightarrow 0$ because $|\gamma_k|^{2m}$ in the summation becomes $1$, so we have truncated our plot for small $k$.

\subsection{Small scale limit}
\begin{figure}[t]
\begin{center}
\vspace{-1cm}
\begin{minipage}{8.1cm}
\includegraphics[height=6cm]{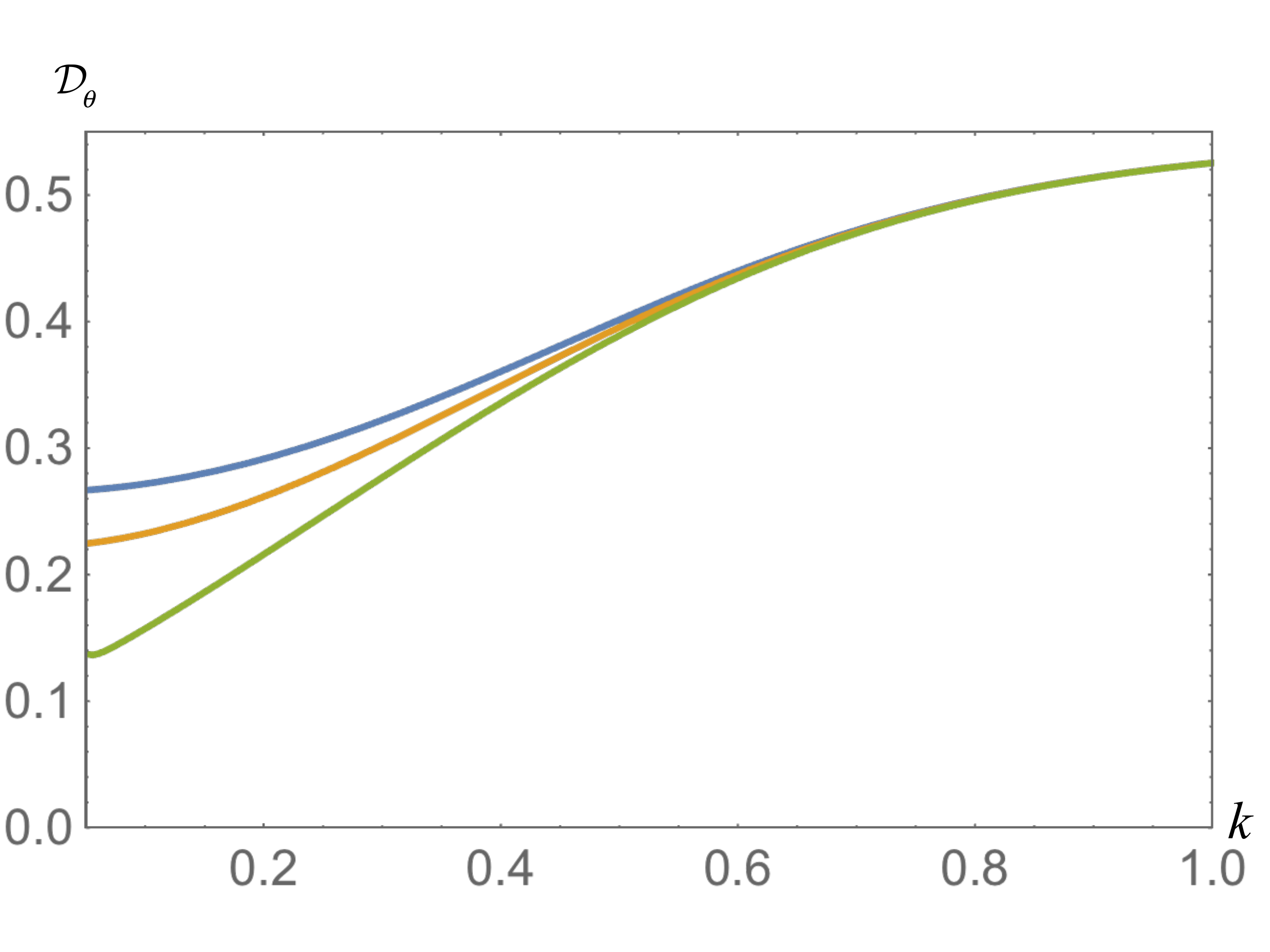}\centering
\end{minipage}
\begin{minipage}{8.1cm}
\hspace{0.5cm}
\includegraphics[height=6cm]{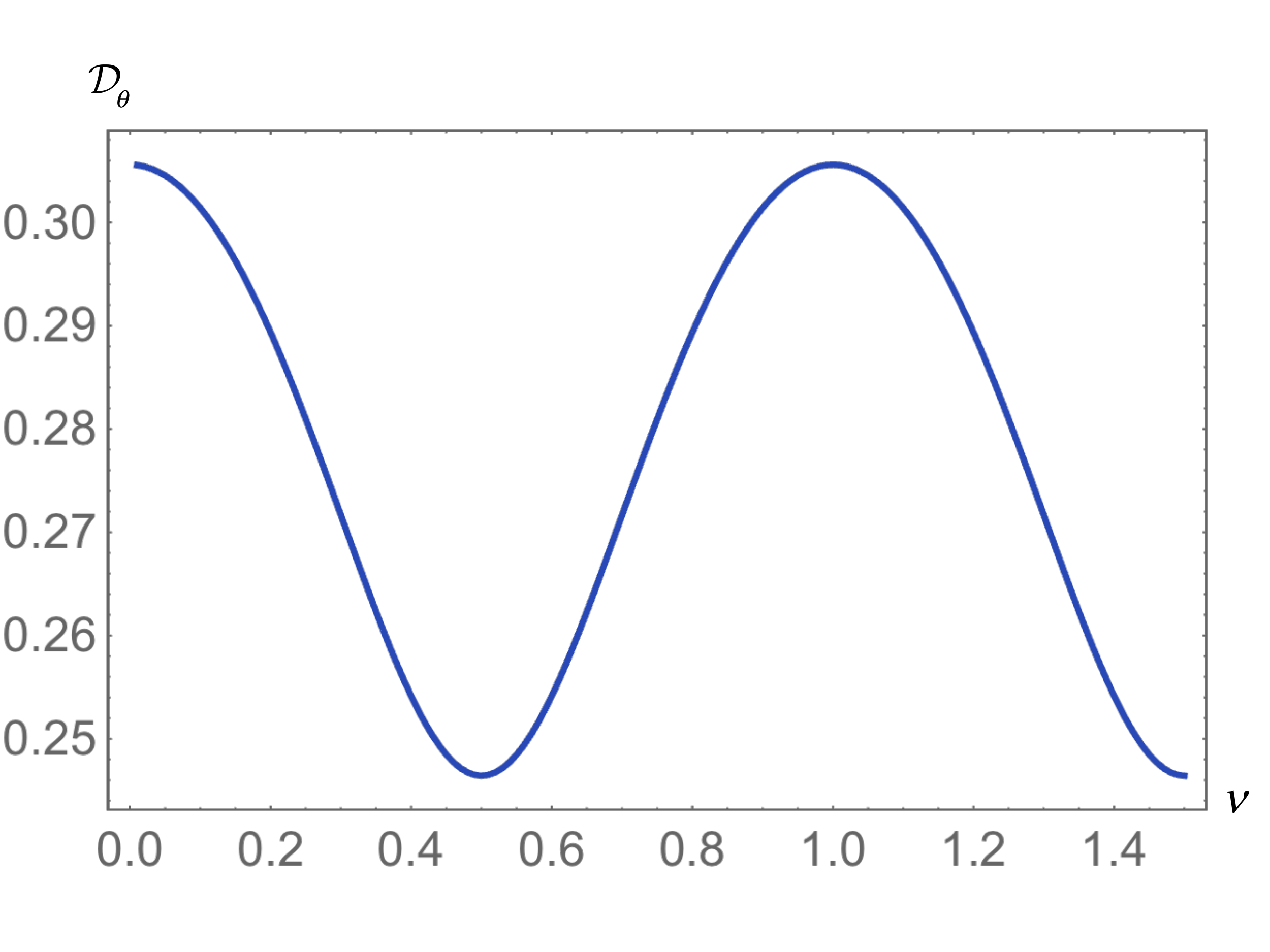}
\end{minipage}
\caption{Plots of the quantum discord as a function of $k$. The Blue line is for $\nu=0$, $1$, the yellow is for $\nu=1/4$, $5/4$ and the green is for $\nu=1/2$, $3/2$.(Left) The quantum discord as a function of $\nu$ when $k=0.25$ and $\theta=\pi/2$. (Right)}
\label{fig5}
\end{center}
\end{figure}

We can calculate the quantum discord (\ref{discord2}) analytically in the $k\rightarrow\infty$ limit for any mass of the scalar field.
Since $\gamma_k\xrightarrow{k\rightarrow\infty}0$, the density matrix $\rho_{A,B}$ becomes
\begin{eqnarray}
\rho_{A,B}&\xrightarrow{k\rightarrow\infty}&\frac{1}{2}\sum_{m=0}^\infty|\gamma_k|^{2m}\left[|0m\rangle\langle0m|
+\frac{m+1}{2}|1m\rangle\langle 1m|
+\sqrt{\frac{m+1}{2}}\,|0m\rangle\langle 1\,m+1|\right.\nonumber\\
&&\left.+\sqrt{\frac{m+1}{2}}\,|1\,m+1\rangle\langle 0m|
+\frac{m+1}{2}\,|1\,m+1\rangle\langle 1\,m+1|\,
\right]\nonumber\\
&\sim&\frac{1}{2}\,|00\rangle\langle 00|+\frac{1}{4}\,|10\rangle\langle 10|
+\frac{1}{2\sqrt{2}}\,|00\rangle\langle 11|+\frac{1}{2\sqrt{2}}\,|11\rangle\langle 00|
+\frac{1}{4}\,|11\rangle\langle 11|\,.
\end{eqnarray}
This can be factorized as
\begin{eqnarray}
\rho_{A,B}&=&\frac{3}{4}\left(\sqrt{\frac{2}{3}}\,|00\rangle+\frac{1}{\sqrt{3}}\,|11\rangle\right)\left(\sqrt{\frac{2}{3}}\,\langle 00|+\frac{1}{\sqrt{3}}\,\langle11|
\right)+\frac{1}{4}\,|10\rangle\langle 10|\,,
\end{eqnarray}
which has eigenvalues of $\rho_{A,B}$ of $3/4$ and $1/4$. On the other hand, for $\rho_{B|\pm}$, we find
\begin{eqnarray}
\rho_{B|\pm}&\xrightarrow{k\rightarrow\infty}&
\frac{1}{2}\sum_{m=0}^\infty|\gamma_k|^{2m}\Bigl[
\left(1\pm\cos\theta\right)|m\rangle\langle m|\Bigr.\nonumber\\
&&\Bigl.+\left(1\mp\cos\theta\right)\Bigl\{
\frac{m+1}{2}\,|m\rangle\langle m|+\frac{m+1}{2}\,|m+1\rangle\langle m+1|
\Bigl\}\Bigl.\nonumber\\
&&\Bigr.
\pm\sin\theta \sqrt{\frac{m+1}{2}}\,|m\rangle\langle m+1|
\pm\sin\theta \sqrt{\frac{m+1}{2}}\,|m+1\rangle\langle m|
\Bigr]\nonumber\\
&\sim&\frac{1}{2}\left(
\begin{array}{cc}
\frac{3}{2}\pm\frac{\cos\theta}{2} & \pm\frac{\sin\theta}{\sqrt{2}}\\
\pm\frac{\sin\theta}{\sqrt{2}} & \frac{1}{2}\mp\frac{\cos\theta}{2}
\end{array}
\right)\,.
\end{eqnarray}
The eigenvalues of $\rho_{B|\pm}$ are then found to be $\frac{1}{2}\left[1\pm\sqrt{\frac{3}{4}\pm\frac{\cos\theta}{2}-\frac{\cos^2\theta}{4}}\right]$. Finally the quantum discord (\ref{discord2}) in this limit is found to be
\begin{eqnarray}
{\cal D}_\theta&=&1+\frac{1}{4}\log_2\frac{1}{4}+\frac{3}{4}\log_2\frac{3}{4}\\
&&
-\frac{1}{2}\left[\,\frac{1}{2}\left(1+\sqrt{\frac{3}{4}+\frac{1}{2}\cos\theta-\frac{1}{4}\cos^2\theta}\right)\log_2\left\{\frac{1}{2}\left(1+\sqrt{\frac{3}{4}+\frac{1}{2}\cos\theta-\frac{1}{4}\cos^2\theta}\,\right)\right\}\right.\nonumber\\
&&\left.\qquad
+\frac{1}{2}\left(1-\sqrt{\frac{3}{4}+\frac{1}{2}\cos\theta
-\frac{1}{4}\cos^2\theta}\right)\log_2\left\{\frac{1}{2}\left(1-\sqrt{\frac{3}{4}+\frac{1}{2}\cos\theta-\frac{1}{4}\cos^2\theta}\,\right)\right\}\right.\nonumber\\
&&\left.
\qquad
+\frac{1}{2}\left(1+\sqrt{\frac{3}{4}-\frac{1}{2}\cos\theta
-\frac{1}{4}\cos^2\theta}\right)\log_2\left\{\frac{1}{2}\left(1+\sqrt{\frac{3}{4}-\frac{1}{2}\cos\theta-\frac{1}{4}\cos^2\theta}\,\right)\right\}\right.\nonumber\\
&&\left.
\qquad
+\frac{1}{2}\left(1-\sqrt{\frac{3}{4}-\frac{1}{2}\cos\theta
-\frac{1}{4}\cos^2\theta}\right)\log_2\left\{\frac{1}{2}\left(1-\sqrt{\frac{3}{4}-\frac{1}{2}\cos\theta-\frac{1}{4}\cos^2\theta}\,\right)\right\}\,
\right]\,.\nonumber
\end{eqnarray}
In this limit, the asymptotic value of the quantum discord ${\cal D}_{\pi/2}$ becomes  $1+\frac{3}{4}\log_2 3 - \frac{\sqrt{3}}{2} \log_2 \left( 2+\sqrt{3} \right) \sim 0.54 $, which agrees with the numerical result in the right panel of Figure~\ref{fig4}. We also plot ${\cal D}_\theta$ as a function of $k$ for different values of the mass parameter $\nu$ in the left panel of Figure~\ref{fig5}. We see that ${\cal D}_\theta$ decreases faster in the cases of $\nu=1/2$ and $3/2$ compared with other masses of the scalar field as we go to large curvatures $k\rightarrow 0$. In order to compare the dependence of $\nu$ for large scales with the entanglement negativity in Figure~\ref{fig3}, we plotted it in the right panel of Figure~\ref{fig5}. The features are similar to one another and the origin of the oscillatory behavior is 
again in the $\cos 2\pi\nu$ in $\gamma_k$ of Eq.~(\ref{gammap}) $(p=k)$.

\section{Summary}
\label{s6}

In this work, we investigated quantum discord between two free modes of a massive scalar field in a maximally entangled state in de Sitter space. We introduced two observers,  one in a global chart and the other in an open chart of de Sitter space, and then  determined the quantum discord created by each detecting one of the modes. This situation is analogous to the relationship between an observer in Minkowski space and another in one of the two Rindler wedges in flat space. In the case of Rindler space, it is known that the entanglement vanishes when the relative acceleration becomes infinite~\cite{FuentesSchuller:2004xp}. In de Sitter space, on the other hand, the observer's relative acceleration corresponds to the scale of the curvature of the open chart.
 We first evaluated entanglement negativity and then quantum discord in de Sitter space. 
We found that the state becomes less entangled as the curvature of the open chart gets larger. In particular, for a massless scalar field $\nu=3/2$ and a conformally coupled scalar field $\nu=1/2$, the entanglement negativity vanishes in the limit of infinite curvature. 
However, we showed that quantum discord does not disappear even in the limit that the entanglement negativity vanishes. In addition, we found that the entanglement negativity survives even in the limit of infinite curvature for a massive scalar field.

In this paper, we considered a simple quantum state (\ref{entangle}). 
It would be interesting to discuss quantum discord in the context of the multiverse~\cite{Kanno:2015lja} (see also related works
\cite{Kanno:2015ewa,Albrecht:2014aga,Collins:2016ahj,Bolis:2016vas}).
It would also be desirable to find a way to prove if
the large scale structure of our universe and temperature fluctuations of the CMB are originated from quantum fluctuations during the initial inflationary era.
 We will leave these issues for future work.

\section*{Acknowledgments}
SK was supported by IKERBASQUE, the Basque Foundation for Science. JShock is grateful for the National Research Foundation (NRF) of South Africa under grant number 87667.
JSoda was in part supported by MEXT KAKENHI Grant Number 15H05895. 
%

\end{document}